\documentclass[pre,twocolumn,notitlepage,10pt,showpacs]{revtex4-1}
\usepackage{graphicx,epstopdf}
\usepackage{amsmath}
\usepackage{amsfonts}
\usepackage{bbm}
\usepackage{amssymb}
\usepackage[utf8]{inputenc}
\usepackage{geometry}
\usepackage[usenames, dvipsnames]{color}
\usepackage{placeins}
\usepackage{natbib}

\usepackage{color}
\usepackage{latexsym}
\usepackage{times,txfonts}

\usepackage{indentfirst}
\usepackage{physics}
\usepackage{hyperref}
\hypersetup{
    colorlinks=true,
    linkcolor=blue,
    filecolor=magenta,      
    urlcolor=cyan,
}

\newcommand{\1}{\mathbbm{1}}

\newcommand{\Nbar}{\overline N}

\begin{document}

\title{Dynamics of the quantum search and quench-induced first-order phase transitions }

\author{Ivan B. Coulamy}
\email{ivanbc@if.uff.br}
\affiliation{Instituto de F\'{i}sica, Universidade Federal Fluminense, Av. Gal. Milton Tavares de Souza s/n, Gragoat\'{a}, 24210-346 Niter\'{o}i, Rio de Janeiro, Brazil}
\author{Andreia Saguia}
\email{amen@if.uff.br}
\affiliation{Instituto de F\'{i}sica, Universidade Federal Fluminense, Av. Gal. Milton Tavares de Souza s/n, Gragoat\'{a}, 24210-346 Niter\'{o}i, Rio de Janeiro, Brazil}
\author{Marcelo S. Sarandy}
\email{msarandy@id.uff.br}
\affiliation{Instituto de F\'{i}sica, Universidade Federal Fluminense, Av. Gal. Milton Tavares de Souza s/n, Gragoat\'{a}, 24210-346 Niter\'{o}i, Rio de Janeiro, Brazil}
\begin{abstract}
We investigate the excitation dynamics at a first-order quantum phase transition (QPT). More specifically, we 
consider the quench-induced QPT in the quantum search algorithm, which aims at finding out a 
marked element in an unstructured list. We begin by deriving the exact dynamics of the model, which is shown 
to obey a Riccati differential equation. Then, we discuss the probabilities of success by adopting 
either global or local adiabaticity strategies. Moreover, we determine the disturbance of the quantum criticality  
as a function of the system size. In particular, we show that the critical point exponentially converges to its 
thermodynamic limit even in a fast evolution regime, which is characterized by both entanglement QPT 
estimators and the Schmidt gap. The excitation pattern is manifested in terms of quantum domains walls 
separated by kinks. The kink density is then shown to follow an exponential scaling as a function of the evolution 
speed, which can be interpreted as a Kibble-Zurek mechanism for first-order QPTs.    
\end{abstract}


\maketitle

\vspace{-0.5 cm}

\section{Introduction}
The implementation of quantum technologies is fundamentally based on a precise control of 
quantum systems. This requires the ability of keeping track of the quantum dynamics 
along a desired path in Hilbert space. In this direction, a successful strategy is 
provided by the adiabatic theorem of quantum mechanics~\cite{Born:28,Kato:50,Messiah:book}. 
It states that a system that is initially prepared in an eigenstate of a time-dependent Hamiltonian 
$H(t)$ will evolve to the corresponding instantaneous eigenstate at a later time $T$, provided that 
$H(t)$ varies smoothly and that $T$ is much larger than (some power of) the relevant minimal inverse 
energy gap (see, e.g., Ref.~\cite{Teufel:03,Jansen:07,Sarandy:04}). The adiabatic theorem is the 
basis for the paradigm of adiabatic quantum computation (AQC)~\cite{Farhi:01}.  Adiabatic 
optimization has been currently implemented and commercially manufactured through quantum 
annealing (QA) devices~\cite{Johnson:11,Berkley:13}, which are based on 
quantum tunnelling due to interactions with a low temperature bath~\cite{Das:08}. 
Such QA devices constitute a promising approach for quantum information processing 
(see, e.g. Refs.~\cite{Boixo:13,Boixo:14,Ronnow:14,Barends:16}).

In AQC, the ground 
state of $H(t)$ adiabatically evolves from an initial simple state to a final state containing the solution of the problem. 
If the process is performed slowly enough, the adiabatic theorem ensures that the 
system stays close to the ground state of  $H(t)$ throughout the evolution. At the final time $T$, measuring the state will give the 
solution of the original problem with high probability. However, the presence of a quantum phase transition (QPT)~\cite{Sachdev:book}  
will imply in the slowdown of the adiabatic evolution, leading to the appearance 
of excitations during the quantum dynamics. This phenomenon is well described by the 
Kibble-Zurek mechanism (KZM)~\cite{KZM:Kibble,KZM:Zurek}. In the quantum realm, a cornerstone lattice 
model in statistical mechanics illustrating the KZM is the transverse-field Ising 
spin-1/2 chain~\cite{Zurek:05,Dziarmaga:05,Polkovnikov:05}. 
In such a case, the ramping from the paramagnetic regime to the ferromagnetic ordering  
does not asymptotically end up in a fully ferromagnetic state. Instead, 
the system will be described by a mosaic of ordered domains 
whose finite size depends on the rate of the transition.  In particular, in the case of  
second-order QPTs, KZM predicts that the size of the ordered domains scales with the transition 
time as a universal power law, which is provided in terms of a combination of  
critical exponents. This approach also reveals many-body 
critical features close to QPTs through the dynamics of the entanglement spectrum~\cite{Canovi:14,Torlai:14}.  

The aim of this work is to investigate the excitation dynamics at a first-order QPT. 
In this direction, we will consider the quench-induced QPT in the quantum search 
Hamiltonian~\cite{Dam:01,Roland:02}, which implements a quantum algorithm 
whose target is to find out a marked element in an unstructured list~\cite{Grover:97}.  
As a first contribution, we will provide the exact dynamics of the model in terms of a single 
Riccati differential equation~\cite{Riccati:Book}. We will then apply this exact solution in 
the characterization of the first-order QPT as well as its associated excitation dynamics. 
For a first-order classical phase transition, KZM has been recently considered in the specific 
case of the two-dimensional Potts model~\cite{KZM-1QPT}. In that case, it has been shown that 
an important role is played by the boundary conditions adopted, which imply into different scaling laws 
for the ordered domains.  The search Hamiltonian is translationally invariant, which leads to scaling 
laws that will be shown to be compatible with those for periodic boundary conditions appearing in 
the classical case. In particular, we will also discuss the probabilities of success of determining the marked 
element along the quantum evolution by adopting either global or local adiabaticity strategies. Moreover, 
we will determine the disturbance of the quantum criticality  as a function of the system size.  
We will then show that the critical point exponentially converges  to its thermodynamic limit even in a fast 
evolution regime. This will be characterized by both entanglement QPT estimators~\cite{Xavier:11} 
(see also Ref.~\cite{Saguia:13}) and the Schmidt gap~\cite{Chiara:12}. As in the transverse-field Ising spin-1/2 chain, 
the excitation pattern will be manifested in terms of quantum domains separated by kinks. However, instead of 
a power law, the kink density will then be shown to follow an exponential scaling as a function of the 
evolution speed, which can be interpreted as a KZM for first-order QPTs.

\section{Dynamics of the quantum search}
The search problem aims at finding out a marked element in an unstructured list of $N$ items. 
In a quantum setting, it can be solved with scaling $\sqrt{N}$, as proven by Grover~\cite{Grover:97}. 
Here, we consider a Hamiltonian implementation through a quantum system composed of $n$ quantum bits (qubits), 
whose Hilbert space has dimension $N=2^n$. The qubits can be taken here as spin-1/2 degrees of freedom arranged in a chain. 
We denote the computational basis by the set 
$\{|i\rangle\}$, with $0\le i\le N-1$. Without loss of generality, we can assume an oracular model such 
that the marked element is the state $| 0 \rangle$. So the implementation of the quantum search 
can be achieved through the projective Hamiltonian
\begin{equation} \label{eq.hamil.def}
H(s)=f(s) (\1-\ketbra{\psi_0})+g(s) (\1-\ketbra 0), 
\end{equation}
where
$|{\psi_0}\rangle = ({1}/{\sqrt N})\sum_{i=0}^{N-1}| i \rangle$ 
and $s$ denotes the normalized time $s=t/T$ ($0 \le s \le 1$), with $T$ the total time of evolution. 
The Grover search has motivated a number  of small scale experimental realizations in different physical 
architectures~\cite{Chuang:98,Jones:98,Kwiat:00,Ahn:00,Bhattacharya:02,Anwar:04,Brickman:05,Walther:05,DiCarlo:09}.
The adiabatic search algorithm starts in $s=0$ with the quantum system prepared in the uniform superposition provided by 
$|\psi(0)\rangle = |\psi_0 \rangle$. This initial state can be split up in the form 
\begin{equation}
|\psi(0)\rangle = a(0) \ket 0 + p(0) \sum_{i=1}^{N-1} \ket i, 
\label{form-ini}
\end{equation} 
with $a(0)=p(0)=1/\sqrt{N}$. 
The system dynamics is then governed by Schr\"odinger equation which, in terms of the normalized time $s$, can be written as
\begin{equation}
 H(s) |\psi(s)\rangle =  \frac{i}{T} |\psi^\prime(s) \rangle ,
 \label{se}
\end{equation}
with $\hbar =1$ and the {\it prime} symbol denoting derivative with respect to $s$. 
Since the Hamiltonian preserves the form of the initial state given in Eq.~(\ref{form-ini}), with $|0\rangle$ as a privileged state, 
the quantum evolution of $|\psi(0)\rangle$ implies in
\begin{equation}
    |\psi(s)\rangle =  a(s) \ket 0 + p(s) \sum_{i=1}^{N-1} \ket i , 
\label{psi-evolved}
\end{equation}
with $a(s)$ and $p(s)$ to be determined by the solution of Eq.~(\ref{se}). 
In order to solve Schr\"odinger equation, we first notice that, 
by defining $|\psi(s)\rangle \equiv a(s)  |\chi(s)\rangle$, Eq.~(\ref{se}) becomes
\begin{equation}
  \left[ H(s) - \frac{i}{T} \frac{a^\prime(s)}{a(s)} \1 \right] |\chi(s)\rangle =  \frac{i}{T}  |\chi^\prime(s) \rangle   ,
  \label{sechi}
\end{equation}
where
\begin{equation}
    |\chi(s)\rangle =   \ket 0 + k(s) \sum_{i=1}^{N-1} \ket i , 
    \label{chis}
\end{equation}
with $k(s)=p(s)/a(s)$. Now, observe that
\begin{eqnarray}
 H(s) |\chi(s)\rangle = f(s) \Nbar \left[ 1-k(s) \right] \ket 0 \nonumber \\
 + \left[ -\frac{f(s)}{N} (1-k(s)) + g(s) k(s) \right] \sum_{i=1}^{N-1} \ket i ,
 \label{Hchi}
\end{eqnarray}
with $\Nbar = 1 - 1/N$. Then, by inserting Eq.~(\ref{Hchi}) into Eq.~(\ref{sechi}), we obtain 
\begin{eqnarray}
f(s) \Nbar \left[ 1-k(s) \right] -\frac{\alpha(s)}{T} &=& 0 ,\label{se1} \\
-\frac{f(s)}{N} \left[ 1-k(s) \right] +g(s)k(s)  -\frac{\alpha(s)}{T} k(s) &=& \frac{i}{T} k^\prime(s) , \,\,\,\,\,\,\,\, \label{se2}
\end{eqnarray}
with $\alpha(s) = i\, a^\prime(s)/a(s)$. 
From Eq.~(\ref{se1}), we can solve the dynamics for $a(s)$, yielding
\begin{equation}
a(s) = \frac{1}{\sqrt{N}}\exp\left\{ -i T \int_0^s f(s^\prime) \Nbar \left[ 1-k(s^\prime) \right] ds^\prime \right\} .
\label{ades}
\end{equation}
Notice that, in Eq.~(\ref{ades}), we have the exponential of a complex number, since $k(s)$ may in general 
exhibit real and imaginary parts. Indeed, the norm of $a(s)$ varies with $s$, since the algorithm targets on 
maximizing 
this probability amplitude at the end of the evolution.  
We can also use Eq.~(\ref{se1}) to eliminate $\alpha(s)$ in Eq.~(\ref{se2}). It then 
follows that $k(s)$ can be obtained by solving
\begin{eqnarray}
  \frac{i}{T} k^\prime(s) &=& f(s) \Nbar k^2(s) +  \left[g(s)+f(s)-2f(s)\Nbar\right]k(s) \nonumber \\
  &&-f(s)(1-\Nbar), 
\label{riccati}
\end{eqnarray}
which is a Riccati equation, i.e. a first-order ordinary differential equation for $k(s)$ that is quadratic in $k(s)$~\cite{Riccati:Book}. Provided the solution of Eq.~(\ref{riccati}), we are able to exactly describe the dynamics 
of the quantum search for an arbitrary number $n$ of qubits. Eq.~(\ref{riccati}) is rather general, holding for any interpolation 
defined by the functions $f(s)$ and $g(s)$. The choice of such functions affects the energy gap from the 
ground state to the first excited state, which determines the time scale of the algorithm. 

\section{Success Probabilities}

In order to investigate the success probabilities of the adiabatic quantum search via Eq.~(\ref{riccati}), we have to define the interpolation 
scheme for the functions $f(s)$ and $g(s)$. As a first step, let us consider the lowest eigenvalues of the eigenspectrum of $H(s)$, which 
are provided by
\begin{equation}
     E_\pm(s)=  \frac{1\pm\sqrt{1-4f(s)g(s)\Nbar}}{2},
\end{equation}
where $E_{-}(s)$ denotes the ground state energy, while $E_+(s)$ is the energy associated with the first excited state. 
Their corresponding eigenstates read
\begin{equation}
    \ket{E_\pm(s)}= {\cal{N}}_\pm(s) \left[ \ket 0 + b_\pm(s) \sum_{i=1}^{N-1} \ket i \right],
\end{equation}
where
\begin{equation}
     b_\pm(s)=  1-\frac{E_\pm(s)}{\Nbar f(s)},
\end{equation}
and ${\cal{N}}_\pm(s)= 1/\sqrt{1+(N-1)b_{\pm}(s)^2}$.
This implies into a gap given by
\begin{equation}
\Delta E(s)=E_+(s)-E_-(s)=  \sqrt{1-4f(s)g(s)\Nbar}.
\end{equation}
In order to stay close to the ground state of $H(s)$, we will impose the adiabatic condition~\cite{Messiah:book}
\begin{equation}
    T \gg \max_s \frac{D(s)}{\Delta E^2(s)},
\label{adco}
\end{equation}
where $D (s) =  \left| \bra  {E_+(s)}  H^\prime(s) \ket {E_-(s)} \right|$. 
In order to evaluate the adiabatic time condition and then analyze the success probabilities of the algorithm, 
we will consider both {\it global} and {\it local} adiabatic strategies. 
In both cases, the system exhibits a first-order QPT at $s_c=1/2$, 
with the energy gap from the ground state to the first excited state 
exponentially shrinking as a function of the input size $n$. This implies 
that, no matter how slowly the system is dynamically driven, its evolution 
cannot follow the  time-dependent ground state close to the quantum 
critical point. More specifically, the system will exhibit excitations 
manifested through the presence of kinks separating domain walls as 
the instantaneous vector state undergoes the QPT.

\subsection{Global adiabatic evolution}

\begin{figure}[!ht]
\centering
\includegraphics[width=0.47\textwidth]{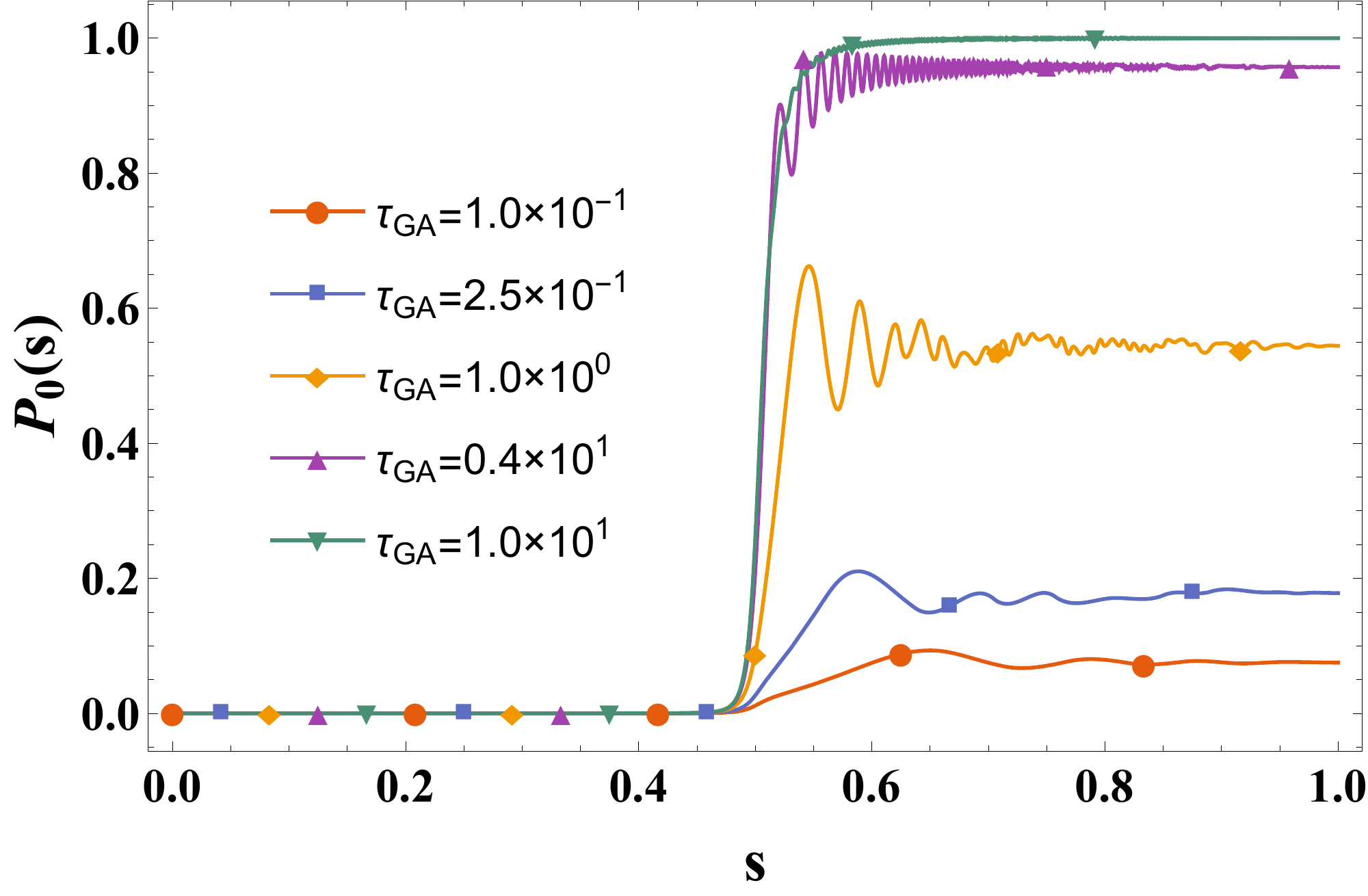}
\caption{Probability of success $P_0(s)$ as a function of the normalized time $s$ for $n=10$ qubits for several dimensionless running rates $\tau_{GA}$, under a global adiabaticity strategy.}
\label{fig:ga2}
\end{figure}

The simplest evolution strategy is to adopt {\it global adiabaticity} through a linear interpolation,  
namely,  
\begin{eqnarray}
f(s)&=&1-s, \nonumber \\
g(s)&=&s. \nonumber
\end{eqnarray}
Therefore, we can directly obtain the running time of the algorithm as $T \gg T_{GA}$, with $T_{GA}$ denoting the 
characteristic time scale for global adiabaticity, which reads
\begin{equation}
    T_{GA}=\max_s \frac{D(s)}{\Delta E^2(s)} = O(N) , 
\end{equation}
with $O(N)$ denoting asymptotic upper bound $N$ on the growth rate of $T_{GA}$. 
Notice then that $T_{GA}$ provides the adiabatic scale for the running time of the algorithm as a function of the size of the 
list. In the particular case of the global adiabaticity strategy, we obtain a linear scaling $N$, which is equivalent to the 
expected scaling in a classical search approach~\cite{Nielsen:book}. 
In the quantum setting, we can now analyze the probability os success $P_0(s)=|\hspace{-0.05cm}\braket{0}{\psi(s)}|^2$
as a function of time. The results are displayed in Fig.~\ref{fig:ga2}, where we consider the dimensionless running ratio 
$\tau_{GA} = T/T_{GA}$ as a measure of adiabaticity. For fast evolutions compared to $T_{GA}$, the adiabatic theorem 
is far from satisfied, which implies into a low probability of success $P_0(s)$. On the other hand, $P_0(s)$ improves as the 
total time gets much greater than $T_{GA}$. Notice also that strong oscillations occur close to the critical point $s_c=1/2$, 
which are reduced at the end of the evolution. This is a consequence of the stiffness of the ordinary differential equation (ODE) 
system~\citep{SolvingODE:book}.  

\subsection{Local adiabatic evolution}
We can improve the time scaling by imposing a {\it local adiabaticity} strategy~\cite{Dam:01,Roland:02}, i.e. 
by dividing the total time into infinitesimal time intervals and applying the adiabaticity condition given by Eq.~(\ref{adco}) 
locally to each of these intervals. By using this procedure, it can be shown that the runtime is minimized for the path (see, e.g. Ref.~\cite{,Kieferova:14})
\begin{eqnarray} 
f(s)&=&1-g(s), \nonumber \\
g(s) &=& \frac{\sqrt{N-1} - \tan\left[\arctan\left(\sqrt{N-1}\right)\left(1-2s\right)\right]}{2\sqrt{N-1}}. \nonumber
\end{eqnarray}
This results in a quadratic speedup over the classical search, i.e., we obtain the 
time complexity $T_{LA} = O(\sqrt{N})$ expected by the Grover quantum search~\cite{Dam:01,Roland:02}. 

\begin{figure}[!ht]  
\includegraphics[width=0.47\textwidth]{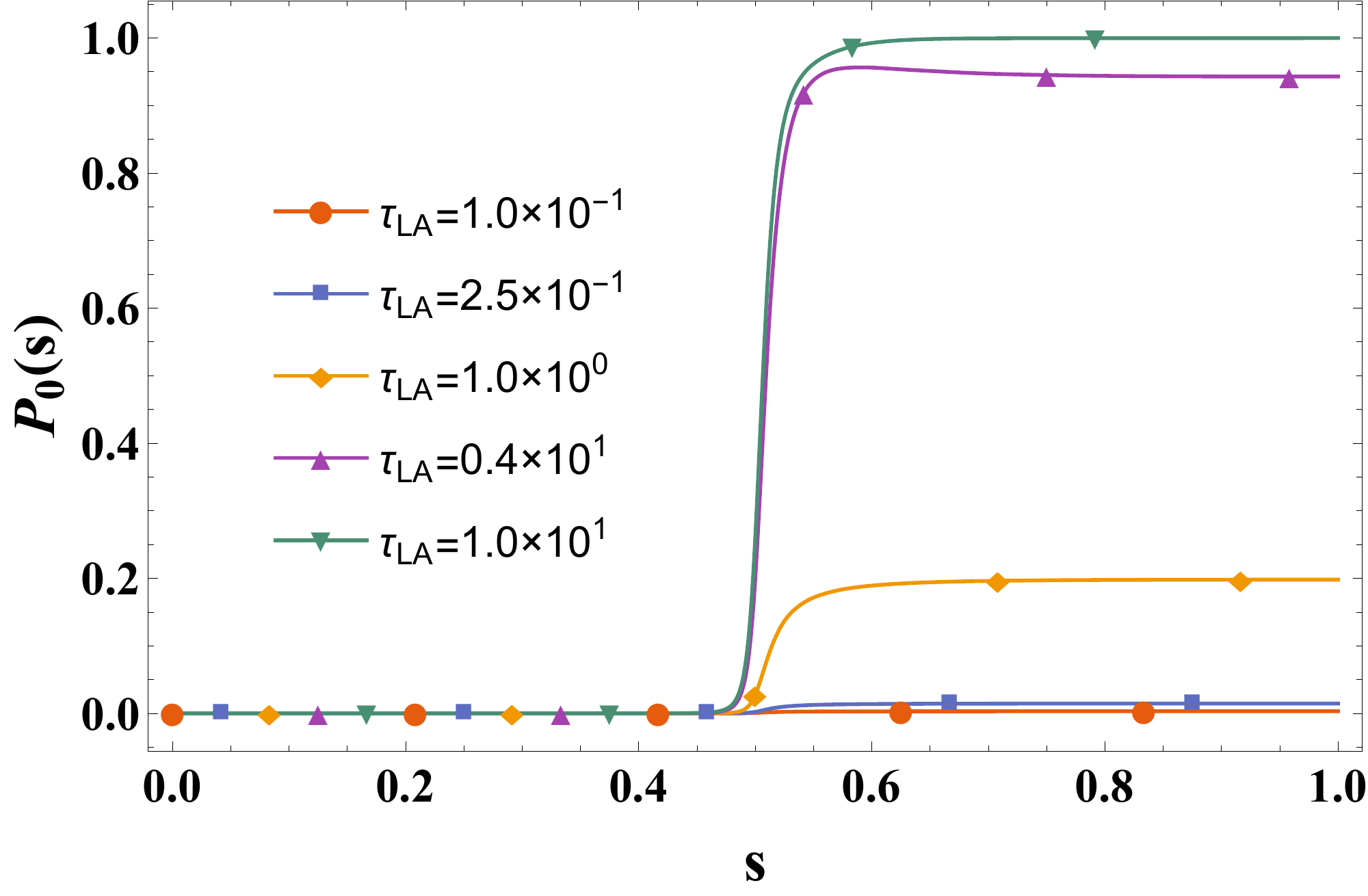}
\caption{Probability of success $P_0(s)$ as a function of the 
normalized time $s$ for n=10 qubits 
for several dimensionless running rates $\tau_{LA}$, under a local adiabaticity strategy.}
\label{fig:la2}
\end{figure}

We can now analyze the probability os success $P_0(s)=|\hspace{-0.05cm}\braket{0}{\psi(s)}|^2$ 
as a function of time for the local strategy. 
The results are displayed in Fig.~\ref{fig:la2}, where we consider the dimensionless running ratio 
$\tau_{LA} = T/T_{LA}$ as a measure of adiabaticity. Notice that the local adiabatic dynamics are more stable, with 
the success probability converging faster after the critical point $s=1/2$ to its final value at $s=1$. Bearing in mind the 
improved asymptotic scaling $O(\sqrt{N})$ of the local adiabaticity strategy, the absence of stiffness in the ODE system, 
and the smoothness of its probability of success as a function of $s$, 
we will adopt this interpolation in the subsequent analysis of the QPT dynamics and quantum domains formation.

\section{Quench-induced first-order QPT}

\subsection{QPT Estimator}
\begin{figure}[!ht]
    \includegraphics[width=0.47\textwidth]{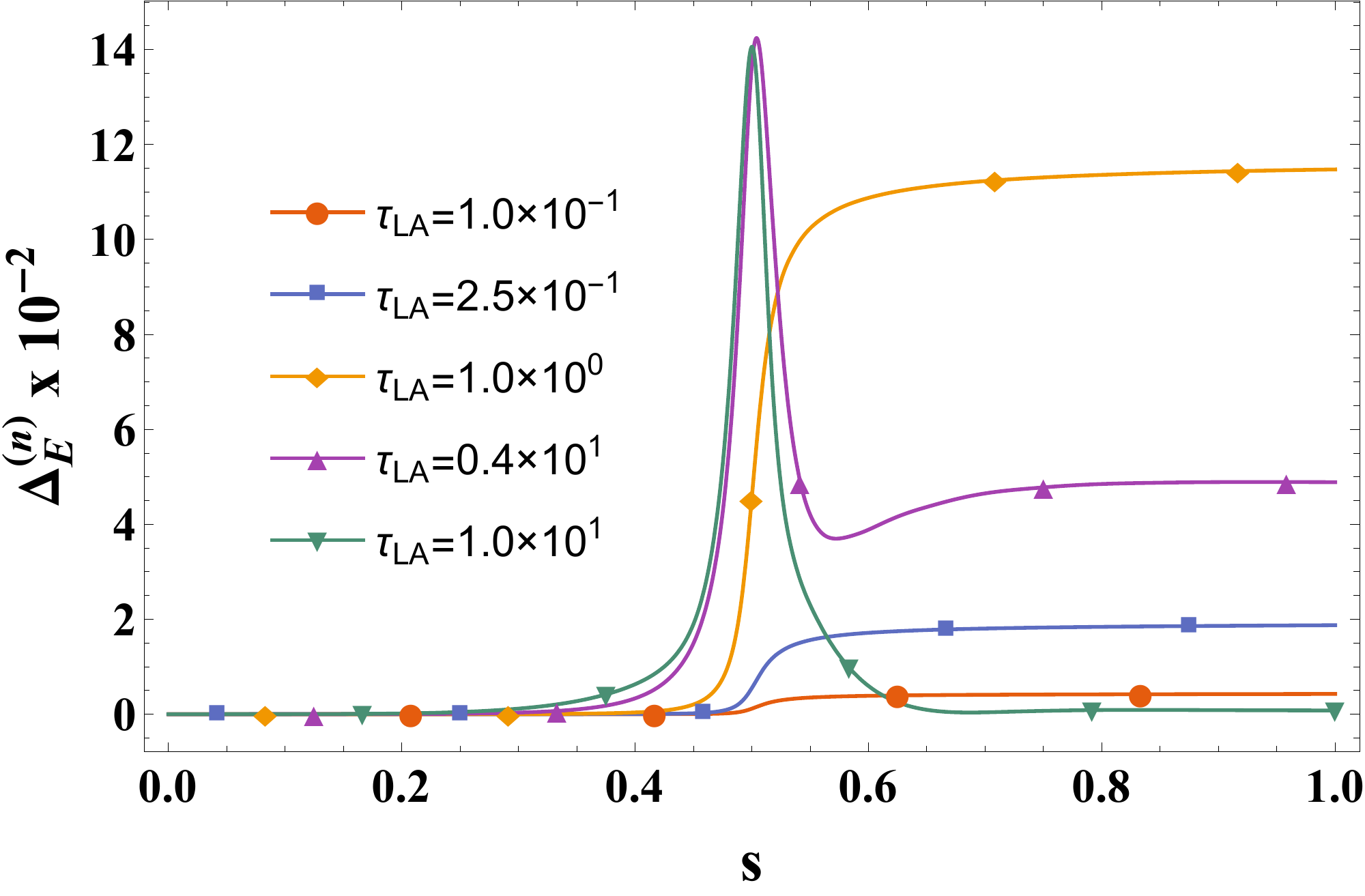}
    \caption{QPT estimator $\Delta_E^{(n)}(s)$ for $n=8$ under local adiabatic evolution for several dimensionless running rates $\tau_{LA}$.}
    \label{fig:est2}
\end{figure}

The characterization of quantum criticality via entanglement estimators~\cite{Xavier:11,Saguia:13} 
is based on the detection of quantum critical points by exploring the distinct behavior of the entanglement entropy in critical and noncritical systems. To begin with, we consider the instantaneous evolved state $|\psi(s)\rangle$  as 
given by Eq.~(\ref{psi-evolved}). By defining a bipartition $AB$ in the quantum system, the density operator of the composite system can be 
written as $\rho_{AB}(s)=|\psi(s)\rangle\langle \psi(s)|$. Then, the entanglement entropy for the subsystem $A$ reads
\begin{equation}
    E(\rho_A)=-\sum_i \lambda_i[\rho_A] \log(\lambda_i[\rho_A]),
\end{equation}
where $\lambda_i[\rho_A]$ denotes the eigenvalues of the reduced density operator $\rho_A = \tr_B \rho_{AB}$.
The entropy $E(\rho_A)$ itself could, in principle, be used to characterize the quantum criticality. However, it usually requires much larger lattices to achieve the same precision as compared with the QPT estimator approach \cite{Xavier:11}. In this scenario, we consider the difference between entanglement entropies for two subsystems with different sizes. Here, we will choose continuous blocks of qubits with sizes $n/2$ and $n/4$. Then, the QPT estimator $\Delta_E^{(n)}(s)$  is defined as
\begin{equation}
    \Delta_E^{(n)}(s)=E(\rho_{n/2})-E(\rho_{n/4}).
\end{equation}
By adopting the local adiabatic interpolation, we provide in Fig.~\ref{fig:est2} the behavior of the quench-induced QPT estimator 
for $n=8$ for several dimensionless times $\tau_{LA}$. As originally observed in Refs.~\cite{Xavier:11,Saguia:14}, 
$\Delta_E^{(n)}(s)$ 
locates a first-order QPT through a peak at the quantum critical point for finite sizes lattices, with the peak tending 
to shrink as the system size is increased. Here, Fig.~\ref{fig:est2} exhibits this peak for $\tau_{LA} > 1$, which 
means a total evolution time $T$ larger than the Grover scaling $O(\sqrt{N})$. For short times $\tau_{LA}$, the peak disappear. Remarkably, the QPT can still be located through the change of concavity in $\Delta_E^{(n)}(s)$. 
\begin{figure}[!ht]
\includegraphics[width=0.47\textwidth]{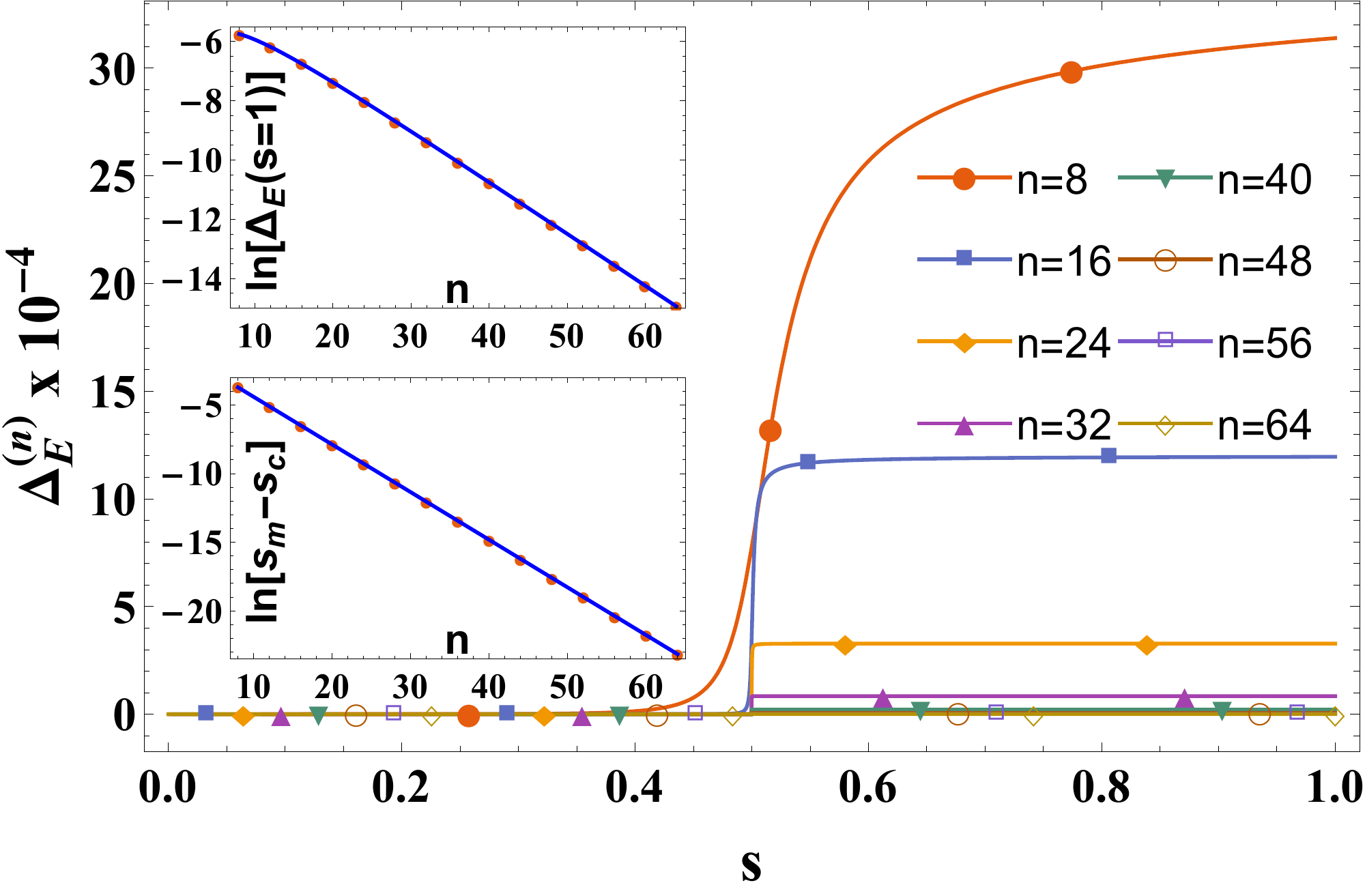}
\caption{Estimator for $n \in \left[ 8 , 64 \right]$ and dimensionless time $\tau_{LA}=0.1$ in the local adiabatic regime as a function of the 
normalized time $s$. The final plateau height at $s=1$ can be fit as $\ln[\Delta_E^{(n)}(s=1)]=-0.18 n -3.5 -6.7/n$. 
The precursor $s_m$ of the critical point $s_{c}$ exponentially converges as $\ln[s_m-s_{c}]=-0.35 n - 0.95$}.
\label{fig:est01n}
\end{figure}

We now analyse the scaling behavior of $\Delta_E^{(n)}(s)$ for different system sizes $n$. We take the local adiabatic strategy in the fast evolution regime. The results are shown in Fig.~\ref{fig:est01n}. Notice that there is a jump in 
$\Delta_E^{(n)}(s)$ as a function of $s$ around the critical point $s_{c}=1/2$, with the plateau after the critical 
point decreasing as the size $n$ gets larger. In the upper inset, we show the plateau height obeys an 
exponential scaling law as a function of $n$. In the lower inset, we show that  that the finite size precursor $s_m$ of the 
critical point exponentially converges to its thermodynamic limit $s_{c}$, with $s_m$ defined as the time $s$ for which 
$\Delta_E^{(n)}(s)$ exhibits an inflection point. This result is remarkable in the sense that the critical point can be 
precisely detected by the QPT estimator even in the fast evolution regime, with exponential convergence of $s_m$ towards the critical point $s_{c}$.   

\subsection{Schmidt Gap}
\begin{figure}[!ht]
    \centering
    \includegraphics[width=0.47\textwidth]{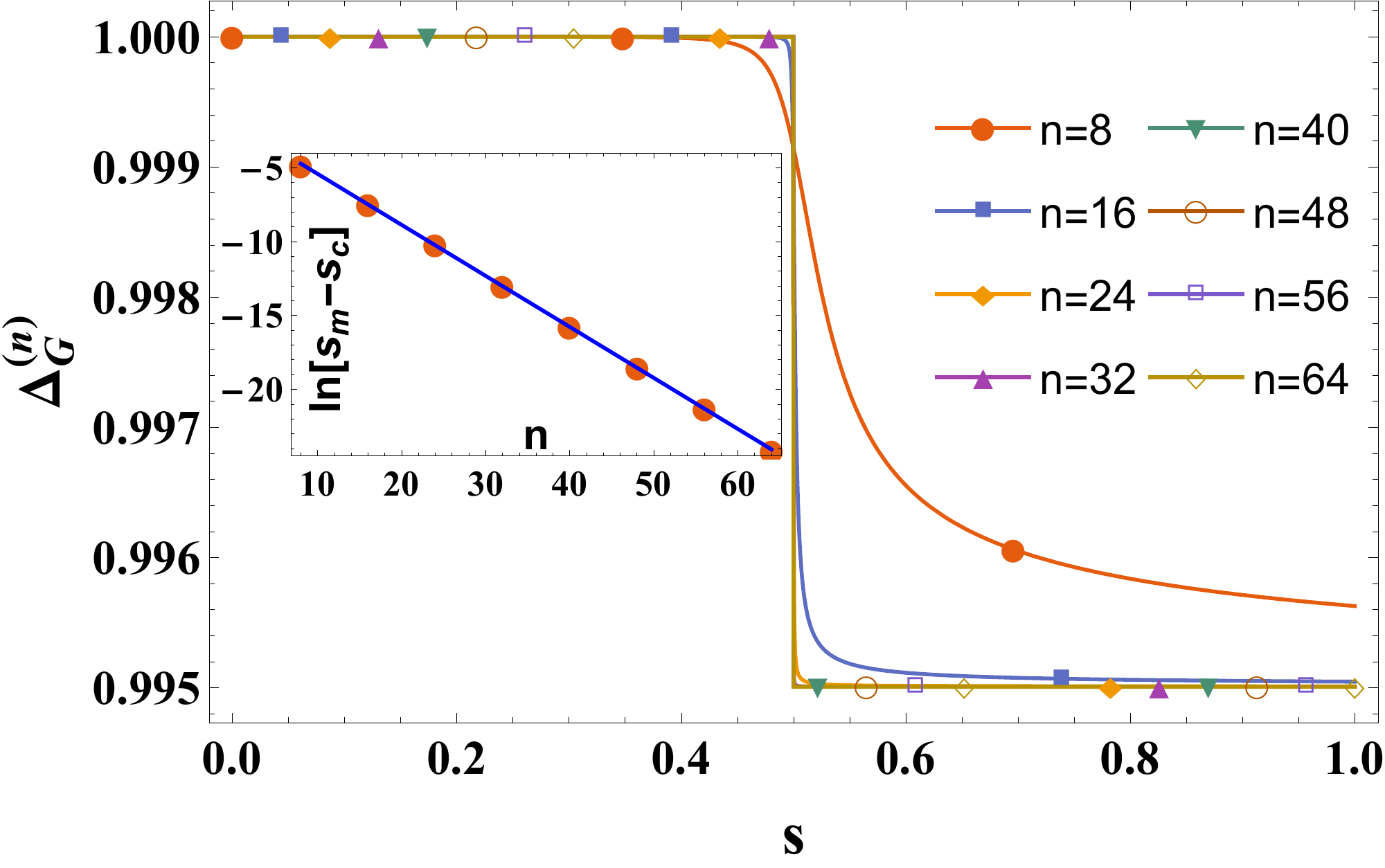}
    \caption{Schmidt gap $\Delta_G^{(n)}(s)$ for $n \in \left[ 8 , 64 \right]$ and dimensionless time $\tau_{LA}=0.1$ in the local adiabatic regime as a function of the normalized time $s$. The precursor $s_m$ of the critical point $s_{c}$ exponentially converges as $\ln[s_m-s_{c}]=-0.34 n - 1.96$}
    \label{fig:schmidt2}
\end{figure}

The Schmidt gap $\Delta_G^{(n)}(s)$ 
is defined as the difference between the two highest eigenvalues of the reduced density matrix 
$\rho_A$ in a composite system $AB$ of $n$ qubits described by the density operator $\rho_{AB}$. 
Here, we will compute the Schmidt gap by splitting up the system into two continuous parts 
with equal size $n/2$ and using the reduced density matrix after tracing out one of the parts. 
When approaching a quantum phase transition, $\Delta_G^{(n)}(s)$ has been shown to signal the critical 
point and to scale with universal critical exponents~\cite{Chiara:12}. For the first-order QPT 
of the quantum search Hamiltonian, we can also show that $\Delta_G^{(n)}(s)$ is able to detect the 
critical point with exponential convergence, as in the case of the QPT estimator. This result is illustrated 
in Fig.~\ref{fig:schmidt2}, where we plot $\Delta_G^{(n)}(s)$ as a function of the normalized time $s$ for several 
system sizes $n$. As $n$ increases, $\Delta_G^{(n)}(s)$ shows a behavior closer to a ladder function. 

\section{Quantum domains and kink dynamics}

In this Section, we will analyze the formation of defects in the quantum search dynamics. 
The initial state for the Hamiltonian in Eq.~(\ref{eq.hamil.def}) is 
$|{\psi(0)}\rangle = ({1}/{\sqrt N})\sum_{i=0}^{N-1}| i \rangle$, which corresponds to a completely polarized state in the Pauli $\sigma_x$ eigenbasis. 
The final expected state, after an ideal adiabatic evolution, is the ferromagnetic state $|{\psi}(s=1)\rangle = | 0 \rangle$, 
which is completely polarized in the Pauli $\sigma_z$ eigenbasis. However, for a quench-induced QPT driven by a finite-time ramping, 
KZM implies into a final state composed by a mosaic of quantum domains separated by kinks. A quantitative discussion has been provided in 
details for a second-order QPT, with the kink density following a typical power-law behavior~\cite{Zurek:05,Dziarmaga:05,Polkovnikov:05}. 
Let us now discuss the kink density behavior for the case of first-order QPTs.  We begin by defining the number 
of kinks through the following observable: 
\begin{equation}
    N_k=1/2 \sum_{i=0}^{n-1}(\1-\sigma^z_i \sigma^z_{i+1}).
    \label{Nop}
\end{equation}
Its expectation value as a function of the normalized time $s$ is then
\begin{equation}
n_k(s) = \langle \psi(s) | N_k | \psi(s) \rangle = \left| a(s) \right|^2 \langle \chi(s) | N_k | \chi(s) \rangle,
\label{nev}
\end{equation}
where we have used that $|\psi(s)\rangle = a(s)  |\chi(s)\rangle$. 
From the normalization of the state vector $|\psi(s)\rangle$ in Eq.~(\ref{psi-evolved}), we obtain
\begin{equation}
\left| a(s) \right|^2 =\frac{1}{1+(N-1)\,\left| k(s) \right|^2} ,
\label{as-norm}
\end{equation}
with $k(s)=p(s)/a(s)$. 
Moreover, by using $ |\chi(s)\rangle$ as given in Eq.~(\ref{chis}), 
we observe that the operator $\sigma^z_i$ will act on the state $|\chi(s)\rangle$ 
by changing the sign of $N/2$ vector elements from $k(s)$ to $-k(s)$. Then
\begin{equation}
    \bra{\chi(s)}  \sigma^z_i \sigma^z_{i+1}  \ket{\chi(s)} =1-\left|k(s)\right|^2.
\label{step-int-kink}
\end{equation}
In order to investigate the domain formation in terms of the evolution speed, we define the density of kinks as 
\begin{equation}
d_k(s) = \frac{1}{n} N_k(s). 
\end{equation}
By using 
Eqs.~(\ref{as-norm}) and (\ref{step-int-kink}) into Eq.~(\ref{nev}), we obtain
\begin{equation}\label{eq.kink}
d_k(s)= \frac{2^{n-1} |k(s)|^2}{1+(2^n-1)|k(s)|^2},
\end{equation} 
Observe that the kink density is completely characterized by the 
amplitude $k(s)$, as given by Eq.~(\ref{eq.kink}). This is highly unusual in 
comparison with the usual KZM. 
It is a consequence of both the 
initial superposition required by the quantum algorithm [as in Eq.~(\ref{form-ini})] 
and the Hamiltonian symmetry, which imposes a uniform superposition 
of all computational states $|i \rangle$ for $i \ne 0$ throughout the evolution 
[as in Eq.~(\ref{psi-evolved})]. Moreover, there is no 
one-to-one association between the energy cost of a domain configuration and the 
excitation density, due to the tower of degenerate excited states arising from the projector 
structure of the Grover Hamiltonian, as given by Eq.~(\ref{eq.hamil.def}). 

\begin{figure}[!ht]
    \centering
    \includegraphics[width=0.47\textwidth]{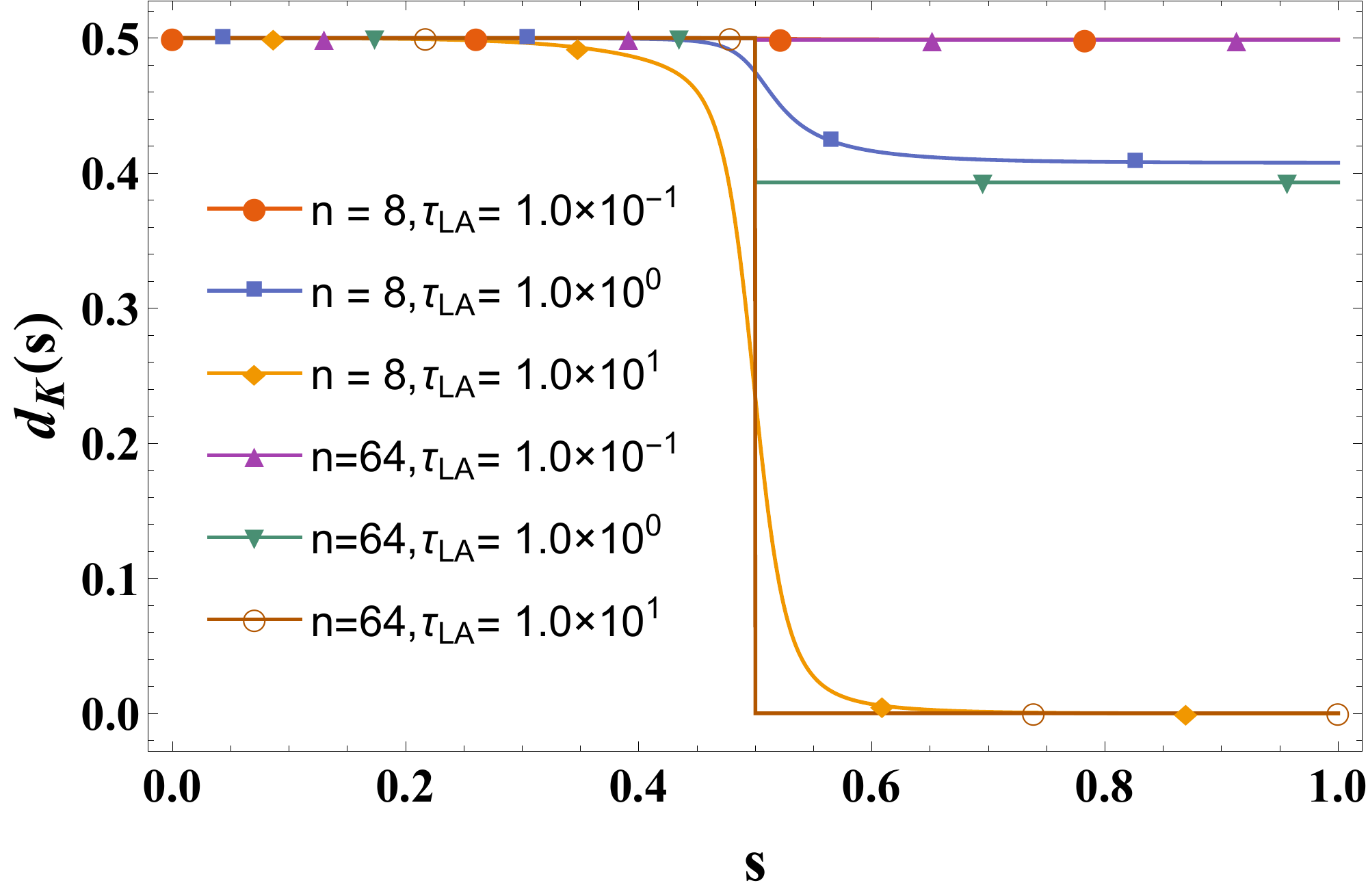}
    \caption{Kink density $d_k(s)$ as a function of the normalized time $s$ for $n=8$ and $n=64$ qubits, where fast and slow speed regimes in terms of the dimensionless time $\tau_{LA}$ are 
    considered.}
    \label{fig:kdyn}
\end{figure}

The behavior of the kink density as a function of the normalized time $s$ for fast and slow 
ramps is illustrated in Fig.~\ref{fig:kdyn}, where the local adiabatic strategy is adopted. Notice 
that, as we increase the dimensionless time $\tau_{LA}$, the kink density tends to decrease 
at $s=1$, yielding a final state closer to the ferromagnetic state. On the other hand, in the fast 
regime, higher excitations are found in the final state, with a kink density closer to its value in 
the original initial state. This result can be already observed for a small lattice such as $n=8$ 
and is shown to hold for larger sizes such as $n=64$ qubits. In particular, the larger the size, 
the closer is the kink density to a ladder function. 
\begin{figure}[!ht]
    \centering
    \includegraphics[width=0.47\textwidth]{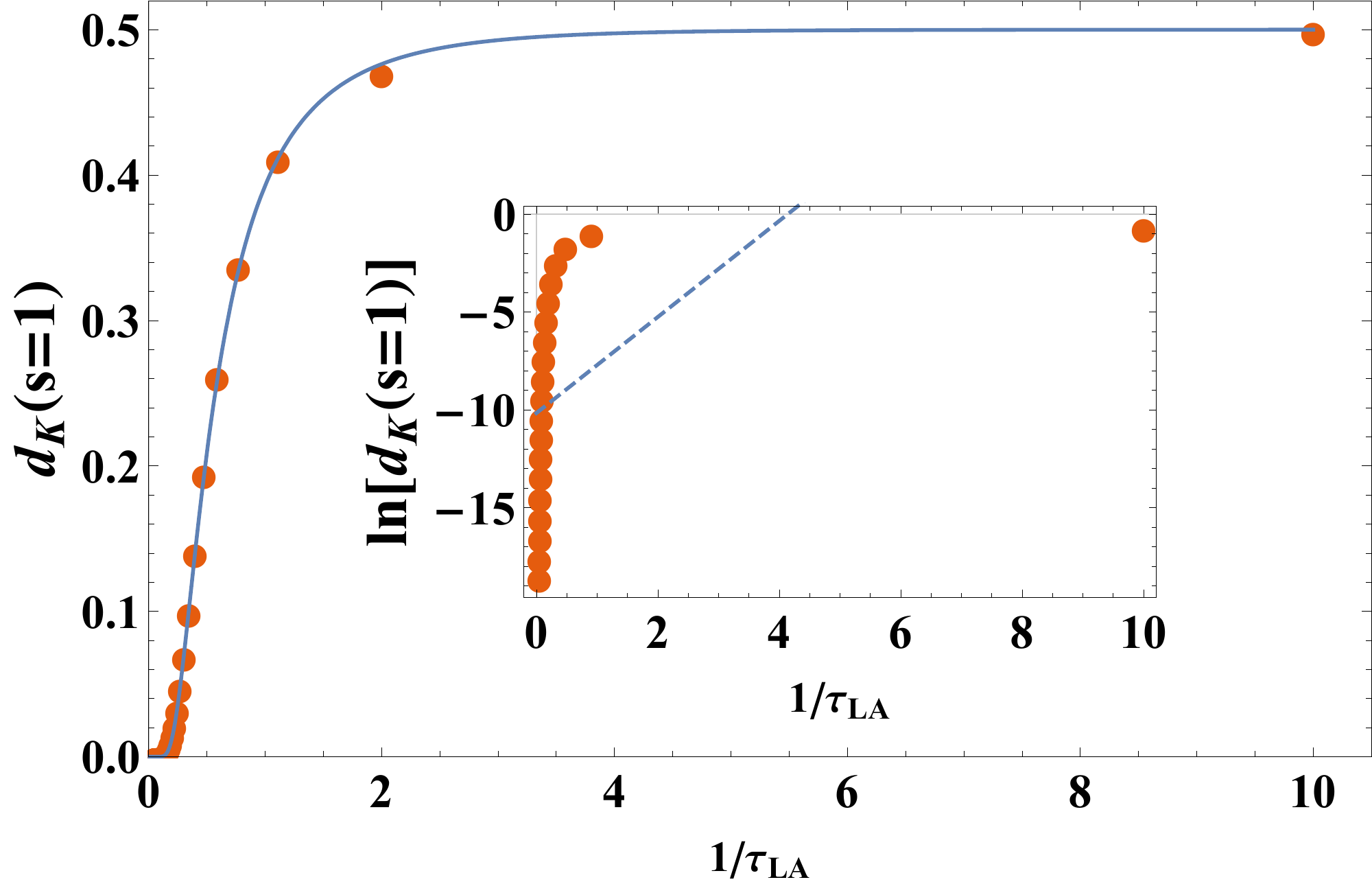}
    \caption{Kink density $d_k$ for $s=1$ as a function of the dimensionless speed $1/\tau_{LA}$ for $n=64$ qubits. The plot can be fit by the curve 
    $d_k = 1/2(1-exp[-a/\tau_{LA}])^{b \tau_{LA}}$, with $a=0.73$ and $b=0.37$. In the inset, we see  a log-plot and its best 
    linear fit, which shows that a power law cannot describe the kink density behavior.}
    \label{fig:kzm}
\end{figure}
In Fig.~\ref{fig:kzm} we consider the kink density as a function of the dimensionless speed $1/\tau_{LA}$ 
for $n=64$ qubits.
As we can see, the kink density obeys an exponential law for its scaling in terms of  $1/\tau_{LA}$. In the inset, we show that the usual 
power law behavior predicted by KZM for second-order QPTs cannot be applied here. Instead, we obtain a KZM for first-order 
QPTs, where the quantum domains appear as expected by a finite speed, but with an exponential scaling law. 
In particular, the convergence of the kink density is 
now much faster than in the case of the traditional KZM, which is due to the exponential behavior of the first-order QPT. 

\section{Conclusion}
We have investigated the dynamics of a first-order QPT through the analysis of the quantum search problem. 
After deriving the exact evolution in terms of a single Riccati equation, we investigated the disturbance of the criticality due to 
the evolution rate. We have shown that the critical point exponentially converges to its thermodynamic limit as a function of the system size.
This scaling law, which is manifested both in QPT entanglement estimators and in the Schmidt gap, holds even in a fast evolution regime. This shows 
that the characterization of the critical point is robust against quench-induced evolutions.  
Remarkably, the QPT estimator does not show a peak in a fast regime, as it is usual for an ideal adiabatic transition.  However, it  
indicates a characterization through an inflection point in the QPT estimator measure. 

Concerning the excitation dynamics, we have derived a 
KZM for first-order QPTs, indicating the existence of an exponential law for the kink density in terms of the dimensionless speed. 
This situation is rather different from the typical (second-order QPT) instances of the KZM, where a polynomial scaling is expected. 
In particular, this implies that the kink density can be used as a useful tool to characterize the order of a quench-induced QPT.  
Moreover, it is related to the disturbance in both the QPT entanglement estimator and Schmidt gap in the dynamical regime. 
Since there in no association between energy cost and domain sizes (due to the tower of degenerate excited states), the 
evolution rate mainly determines the presence or absence of domain walls if a detection scheme (measurement) is performed 
on the system. In any case, for the Grover dynamics, the kink density still reflects an exponential behavior as a function of the evolution rate 
(as illustrated in Fig.~\ref{fig:kzm}). 

The quantum search Hamiltonian is the main representative of a larger 
class of projector-based Hamiltonians, which can be used to implement more general quantum 
algorithms. We expect the pattern of excitation dynamics derived in our 
work applies to these generalized models as well (including the quench-induced QPT behavior).
Naturally, these models are very different from the usual KZM in Ising spin glasses, e.g. 
in Ref.~\cite{Dutta:book}. 
 As a future perspective, we intend to consider decoherence effects~\cite{Dutta:16} in dynamical first-order QPTs. 
 Moreover, we are also interested in the exchange between power-law and exponential-law behaviors as a 
 consequence of boundary conditions~\cite{KZM-1QPT} and Hamiltonian symmetries.

\section*{Acknowledgments}
I.B.C. is supported by CNPq-Brazil. M.S.S. acknowledges support from CNPq-Brazil (No. 304237/2012-4), 
FAPERJ (No 203036/2016), and the Brazilian National Institute for Science and Technology of Quantum Information (INCT-IQ).

\end{document}